\documentclass[fleqn,twoside]{article}
\usepackage{espcrc2}

\usepackage{graphicx}
\usepackage[figuresright]{rotating}

\newcommand{\AmS}{{\protect\the\textfont2
  A\kern-.1667em\lower.5ex\hbox{M}\kern-.125emS}}

\title{Recent results on rare decay $\pi^{0}\rightarrow e^{+}e^{-} $ }

\author{A.E. Dorokhov\address[MCSD]{Joint Institute for Nuclear Research,
Bogoliubov Laboratory of Theoretical
Physics \\
        114980, Moscow region, Dubna, Russia}%
        \thanks{This work is
partially supported by the Russian Foundation for Basic Research
projects No. 08-02-08110-z, the
Scientific School grant 195.2008.2.}}

\begin{document}

\begin{abstract}
Experimental and theoretical progress concerning the rare decay
$\pi^{0}\rightarrow e^{+}e^{-} $ is briefly reviewed. It includes the latest data
from KTeV and a new model independent estimate of the decay branching which show the
deviation between experiment and theory at the level of $3.3\sigma$.
\vspace{1pc}
\end{abstract}

\maketitle

\section{Introduction}

Astrophysics observables tell us that $95\%$ of the matter in the Universe is
not described in terms of the Standard Model (SM) matter. Thus, the search for
the traces of New Physics is a fundamental problem of particle physics. There
are two strategies to look for the effects of New Physics: experiments
at high energy and experiments at low energy. In high-energy experiments
it is considered that due to a huge amount of energy the heavy degrees of freedom presumably
characteristic of the SM extension sector are possible to excite.
In low-energy experiments it is huge statistics
that compensates the lack of energy by measuring the rare processes characteristic
of such extensions. At present, there is no any
evidence for deviation of SM predictions from the results of high-energy
experiments and we are waiting for the LHC epoch. On the other hand, in low-energy experiments
there are rough edges indicating
such deviations. The most famous example is the muon $(g-2)$. Below it will be shown
that due to recent experimental and theoretical progress the rare process
$\pi^{0}\rightarrow e^{+}e^{-} $ became a good SM test process and that at the moment
there is a discrepancy between the SM prediction and experiment at the level of $3.3\sigma$
deviation.

\section{KTeV data}

In 2007, the KTeV collaboration published the result \cite{Abouzaid:2007md} for
the branching ratio of the pion decay into an
electron-positron pair
\begin{equation}
B_{\mathrm{no-rad}}^{\mathrm{KTeV}}\left(  \pi^{0}\rightarrow e^{+}%
e^{-}\right)  =\left(  7.48\pm0.38\right)  \cdot10^{-8}. \label{KTeV}%
\end{equation}
The result is based on observation of 794 candidate $\pi^0 \rightarrow e^+e^-$ events using
$K_L \rightarrow 3\pi^0$ as a source of tagged $\pi^0$s. Due to a complicated chain
of the process and a good technique for final state resolution used by KTeV this is
a process with low background.

\section{Classical theory of $\pi^{0}\rightarrow e^{+}e^{-}$ decay}

The rare decay $\pi^{0}\rightarrow e^{+}e^{-}$ has been studied theoretically
over the years, starting with the first prediction of the rate by Drell
\cite{Drell59}. Since no spinless current coupling of quarks to leptons
exists, the decay is described in the lowest order of QED as a one-loop
process via the two-photon intermediate state, as shown in Fig. 1. A factor of
$2\left(  m_{e}/m_{\pi}\right)  ^{2}$ corresponding to the approximate
helicity conservation of the interaction and two orders of $\alpha$ suppress
the decay with respect to the $\pi^{0}\rightarrow\gamma\gamma$ decay, leading
to an expected branching ratio of about $10^{-7}$. In the Standard Model
contributions from the weak interaction to this process are many orders of
magnitude smaller and can be neglected.


\begin{figure}[th]
\includegraphics[width=5cm]{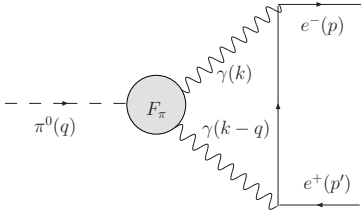}\caption{Triangle diagram for the $\pi
^{0}\rightarrow e^{+}e^{-}$ process with a pion $\pi^{0}\rightarrow
\gamma^{\ast}\gamma^{\ast}$ form factor in the vertex.}%
\label{fig:triangle}%
\end{figure}

To the lowest order in QED the normalized branching ratio is given by%
\begin{eqnarray}
R\left(  \pi^{0}\rightarrow e^{+}e^{-}\right)  =\frac{B\left(  \pi
^{0}\rightarrow e^{+}e^{-}\right)  }{B\left(  \pi^{0}\rightarrow\gamma
\gamma\right)  }\\ \label{B}\
=2\left(  \frac{\alpha}{\pi}\frac{m_{e}}{m_{\pi}}\right)
^{2}\beta_{e}\left(  m_{\pi}^{2}\right)  \left\vert \mathcal{A}\left(  m_{\pi
}^{2}\right)  \right\vert ^{2}, \nonumber%
\end{eqnarray}
where $\beta_{e}\left(  q^{2}\right)  =\sqrt{1-4\frac{m_{e}^{2}}{q^{2}}}$,
$B\left(  \pi^{0}\rightarrow\gamma\gamma\right)  =0.988$.
The amplitude $\mathcal{A}$ can be written as%

\begin{eqnarray}
\mathcal{A}\left(  q^{2}\right)=\frac{2i}{q^{2}}\int\frac{d^{4}k}{\pi^{2}}
F_{\pi\gamma^{\ast}\gamma^{\ast}}\left(  k^{2},\left(  k-q\right)  ^{2}\right) \\ \label{R}
\cdot\frac{q^{2}k^{2}-\left(  qk\right)  ^{2}}
{\left(  k^{2}+i\varepsilon\right)
\left(  \left(  k-q\right)  ^{2}+i\varepsilon\right)  \left(  \left(
k-p\right)  ^{2}-m_{e}^{2}+i\varepsilon\right)},\nonumber
\end{eqnarray}
where $q^{2}=m_{\pi}^{2},p^{2}=m_{e}^{2}$.
$F_{\pi\gamma^{\ast}\gamma^{\ast}}$ is the form
factor of the transition $\pi^{0}\rightarrow\gamma^{\ast}\gamma^{\ast}$ with
off-shell photons.

The imaginary part of $\mathcal{A}$ is defined uniquely as
\begin{eqnarray}
&&\mathrm{Im}\mathcal{A}\left(  q^{2}\right)  =\frac{\pi}{2\beta
_{e}\left(  q^{2}\right)  }\ln\left(  y_{e}\left(  q^{2}\right)  \right)
,\label{Im}\\
&& y_{e}\left(  q^{2}\right)  =\frac{1-\beta_{e}\left(  q^{2}\right)
}{1+\beta_{e}\left(  q^{2}\right)  }.\nonumber%
\end{eqnarray}
It comes from the contribution of real photons in the intermediate state and is
model independent since $F_{\pi\gamma^{\ast}\gamma^{\ast}}\left(  0,0\right)
=1$. Using inequality $\left\vert \mathcal{A}\right\vert ^{2}\geq\left(
\mathrm{Im}\mathcal{A}\right)  ^{2}$
one can get the well-known unitary bound for the branching ratio \cite{Berman60}%
\begin{eqnarray}
B\left(  \pi^{0}\rightarrow e^{+}e^{-}\right) \label{UnitB}\\
\geq B^{\mathrm{unitary}}\left(  \pi^{0}\rightarrow e^{+}e^{-}\right)  =4.69\cdot10^{-8}.
\nonumber\end{eqnarray}

One can attempt to reconstruct the full amplitude by using
a once-subtracted dispersion relation \cite{Bergstrom:1983ay}
\begin{equation}
\mathcal{A}\left(  q^{2}\right)  =\mathcal{A}\left(  q^{2}=0\right)
+\frac{q^{2}}{\pi}\int_{0}^{\infty}ds\frac{\mathrm{Im}\mathcal{A}\left(
s\right)  }{s\left(  s-q^{2}\right)  }.\label{DispRel}%
\end{equation}
If one assumes that Eq. (\ref{Im}) is valid for any $q^2$, then
one arrives for $q^{2}\geq4m_{e}^{2}$ at
\cite{D'Ambrosio:1986ze,Savage:1992ac,Ametller:1993we}%
\begin{eqnarray}
\mathrm{Re}\mathcal{A}\left(  q^{2}\right)=\mathcal{A}\left(
q^{2}=0\right) +\frac{1}{\beta_{e}\left(  q^{2}\right)  }\label{Rqb} \\
\cdot  \left[  \frac{1}%
{4}\ln^{2}\left(  y_{e}\left(  q^{2}\right)  \right)  +\frac{\pi^{2}}%
{12}+\mathrm{Li}_{2}\left(  -y_{e}\left(  q^{2}\right)  \right)  \right],
\nonumber\end{eqnarray}
where $\mathrm{Li}_{2}\left(  z\right)  =-\int_{0}^{z}\left(  dt/t\right)
\ln\left(  1-t\right)  $ is the dilogarithm function.
The second term in Eq.~(\ref{Rqb}) takes into account a strong $q^{2}$
dependence of the amplitude around the point $q^{2}=0$ occurring due to the
branch cut coming from the two-photon intermediate state. In
the leading order in $\left(  m_{e}/m_{\pi}\right)  ^{2},$ Eq. (\ref{Rqb})
reduces to
\begin{equation}
\mathrm{Re}\mathcal{A}\left(  m_{\pi}^{2}\right)  =\mathcal{A}\left(
q^{2}=0\right)  +\ln^{2}\left(  \frac{m_{e}}{m_{\pi}}\right)  +\frac{\pi^{2}%
}{12}.
\end{equation}

Thus, the amplitude is fully reconstructed up to a subtraction constant.
Usually, this constant containing the nontrivial dynamics of the process is
calculated within different models describing the form factor $F_{\pi}%
(k^{2},q^{2})$
\cite{Bergstrom:1982zq,Bergstrom:1983ay,Savage:1992ac,Efimov:1981vh,Dorokhov:2007bd}.
However, it has recently been shown
in \cite{Dorokhov:2007bd} that this constant may be expressed in terms of the inverse
moment of the pion transition form factor given in symmetric kinematics of
spacelike photons%
\begin{eqnarray}
&&\mathcal{A}\left(  q^{2}=0\right)  =3\ln\left(  \frac{m_{e}}{\mu}\right)
-\frac{5}{4}\label{R0}\\
&&-\frac{3}{2}\left[  \int_{0}^{\mu^{2}}dt\frac{F_{\pi\gamma^{\ast}\gamma^{\ast
}}\left(  t,t\right)  -1}{t} \right.\nonumber\\
&&+\left.\int_{\mu^{2}}^{\infty}dt\frac{F_{\pi\gamma
^{\ast}\gamma^{\ast}}\left(  t,t\right)  }{t}\right]  .\nonumber
\end{eqnarray}
Here, $\mu$ is an arbitrary (factorization) scale. One has to note that the
logarithmic dependence of the first term on $\mu$ is compensated by the scale
dependence of the integrals in the brackets. In this way two independent processes becomes
related.

\section{Importance of CLEO data on $F_{\pi\gamma^{\ast}\gamma}$}

In order to estimate the integral in Eq.~(\ref{R0}), one needs to define
the pion transition form factor in symmetric kinematics for spacelike photon
momenta. Since it is unknown from the first principles, we will adapt the
available experimental data to perform such estimates. Let us first use the
fact that $F_{\pi\gamma^{\ast}\gamma^{\ast}}\left(  t,t\right)  <F_{\pi
\gamma^{\ast}\gamma^{\ast}}\left(  t,0\right)  $ for $t>0$ in order to obtain
the lower bound of the integral in Eq.~(\ref{R0}). For this purpose, we take
the experimental results from the CELLO \cite{Behrend:1990sr} and CLEO
\cite{Gronberg:1997fj} Collaborations for the pion transition form factor in
asymmetric kinematics for spacelike photon momentum which is well
parametrized by the monopole form \cite{Gronberg:1997fj}
\begin{eqnarray}
F_{\pi\gamma^{\ast}\gamma^{\ast}}^{\mathrm{CLEO}}\left(  t,0\right)  =\frac
{1}{1+t/s_{0}^{\mathrm{CLEO}}},\label{VMD1}\\
 s_{0}^{\mathrm{CLEO}}=\left(
776\pm22\quad\mathrm{MeV}\right)  ^{2}.\nonumber
\end{eqnarray}

\begin{figure}[th]
\includegraphics[width=8cm]{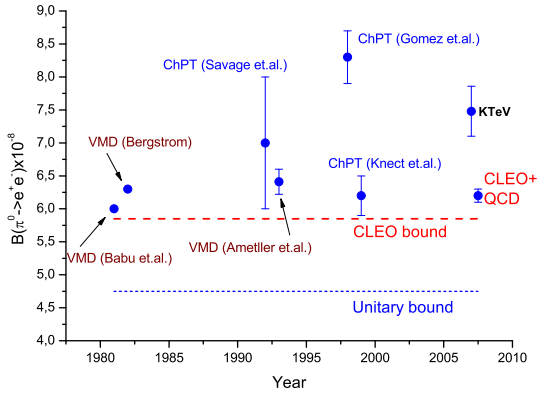}\caption{Evolution
of model predictions and comparison with the latest KTeV result.}%
\label{fig:triangle}%
\end{figure}

For this type of the form factor one finds from Eq.~(\ref{R0})
that
\begin{eqnarray}
\mathcal{A}\left(  q^{2}=0\right) \label{Rcleo}\\
 >-\frac{3}{2}\ln\left(  \frac
{s_{0}^{\mathrm{CLEO}}}{m_{e}^{2}}\right)  -\frac{5}{4}=-23.2\pm
0.1.\nonumber
\end{eqnarray}
Thus, for the branching ratio we are able to establish the important lower
bound which considerably improves the unitary bound given by Eq.~(\ref{UnitB})
\begin{eqnarray}
&&B\left(  \pi^{0}\rightarrow e^{+}e^{-}\right) \label{Bcleo}\\
&& >B^{\mathrm{CLEO}}\left(
\pi^{0}\rightarrow e^{+}e^{-}\right)  =\left(  5.84\pm0.02\right)\cdot10^{-8}.
\nonumber\end{eqnarray}

It is natural to assume that the monopole form is
also a good parametrization for the form factor in symmetric kinematics
\begin{eqnarray}
F_{\pi\gamma^{\ast}\gamma^{\ast}}\left(  t,t\right)  =\frac{1}{1+t/s_{1}}.\label{Ftt}%
\end{eqnarray}
The scale $s_{1}$ can be fixed from the relation for the slopes of the form
factors in symmetric and asymmetric kinematics at low $t$
\cite{Dorokhov:2003sc},
\begin{equation}
\left.  -\frac{\partial F_{\pi\gamma^{\ast}\gamma^{\ast}}\left(  t,t\right)
}{\partial t}\right\vert _{t=0}=\left.  -2\frac{\partial F_{\pi\gamma^{\ast
}\gamma^{\ast}}\left(  t,0\right)  }{\partial t}\right\vert _{t=0}%
,\label{slope}%
\end{equation}
that gives $s_{1}=s_{0}/2$. Note that a similar reduction of the scale is also
predicted by OPE QCD from the large momentum behavior of the form factors:
$s_{1}^{OPE}=s_{0}^{OPE}/3$ \cite{Lepage:1980fj}. Thus, the estimate for
$\mathcal{A}\left(  0\right)$ can be obtained from
Eq.~(\ref{Rcleo}) by shifting the lower bound by a positive number which
belongs to the interval $[3\ln(2)/2,3\ln(3)/2]$
\begin{equation}
\mathcal{A}\left(  q^{2}=0\right)  =-\frac{3}{2}\ln\left(  \frac{s_{1}}%
{m_{e}^{2}}\right)  -\frac{5}{4}=-21.9\pm0.3.\label{R0t}%
\end{equation}
With this result the branching ratio becomes
\begin{equation}
B\left(  \pi^{0}\rightarrow e^{+}e^{-}\right)  =\left(  6.23\pm0.09\right)
\cdot10^{-8}.\label{Bt}%
\end{equation}
This is $3.3$ standard deviations lower than the KTeV result given by
Eq.~(\ref{KTeV}).

\section{Other decay modes}
The $\eta\rightarrow l^{+}l^{-}$ decay can be analyzed in a similar manner.
As in the pion case, the CLEO Collaboration has parametrized the data for the
$\eta$-meson in the monopole form \cite{Gronberg:1997fj}:
\begin{eqnarray}
F_{\eta\gamma^{\ast}\gamma^{\ast}}^{\mathrm{CLEO}}\left(  t,0\right)
=\frac{1}{1+t/s_{0\eta}^{\mathrm{CLEO}}},\\
 s_{0\eta}^{\mathrm{CLEO}}=\left(  774\pm29\quad\mathrm{MeV}\right)  ^{2},
\nonumber\end{eqnarray}
which is very close to the relevant pion parameter. Then following the previous
case (with evident substitutions), one finds the bounds for the $q^{2}%
\rightarrow0$ limit of the amplitude $\eta\rightarrow\mu^{+}\mu^{-}$ as
\begin{eqnarray}
&&\mathcal{A}_{\eta}\left(  q^{2}=0\right)  \\
&&>-\frac{3}{2}\ln\left(
\frac{s_{0\eta}^{\mathrm{CLEO}}}{m_{\mu}^{2}}\right)  -\frac{5}{4}=-\left(
7.2\pm0.1\right)  ,
\nonumber\end{eqnarray}
and for $\eta\rightarrow e^{+}e^{-}$ one gets again Eq.~(\ref{Rcleo}). The
obtained estimates allow one to find the bounds for the branching ratios
\begin{eqnarray}
B\left(  \eta\rightarrow\mu^{+}\mu^{-}\right)     <\left(  6.23\pm
0.12\right)  \cdot10^{-6},\label{RbEta}\\
B\left(  \eta\rightarrow e^{+}e^{-}\right)     >\left(  4.33\pm0.02\right)
\cdot10^{-9}.\nonumber
\end{eqnarray}
It is important to note that for the decay $\eta\rightarrow\mu^{+}\mu^{-}$
we get the upper limit
for the branching. This is because the real part of the amplitude for this
process taken at the physical point $q^{2}=m_{\eta}^{2}$ for the parameter
$s_{0\eta}^{\mathrm{CLEO}}$ remains negative and a positive shift due to the
change of the scale $s_{0\eta}\rightarrow s_{1\eta}$ reduces the absolute
value of the real part of the amplitude $\left\vert \mathrm{Re}%
\mathcal{A}\left(  m_{\eta}^{2}\right)  \right\vert $. At the same time,
considering the decays of $\pi^{0}$ and $\eta$ into an electron-positron pair,
the evolution to physical point (\ref{Rqb}) makes the real part of the
amplitude to be positive for the parameter $s_{0}^{\mathrm{CLEO}}$ and the
absolute value of the real part of the amplitude increases in changing the
scales of the meson form factors. Thus, it would be very interesting to check
experimentally the predicted bounds for the process
$\eta\rightarrow\mu^{+}\mu^{-}$.

The predictions for the decays
$\eta\rightarrow l^{+}l^{-}$ obtained by reducing the scale $s_{0\eta
}^{\mathrm{CLEO}}\rightarrow s_{1\eta}$ for the case of the $\eta$-meson
transition form factor are given in Table \ref{table2}.
Also note the recent measurement by the WASA/Celcius Collaboration
\cite{Berlowski:2008zz} which improves
the upper limit for the branching $\eta\rightarrow e^{+}e^{-}$.

\begin{table*}[th]
\caption[Results]{Values of the branchings $B\left(  P\rightarrow l^{+}%
l^{-}\right)  $ obtained in our approach and compared with the available
experimental results. }%
\begin{tabular}
[c]{|c|c|c|c|c|}\hline
$B$ & Unitary bound & CLEO bound & CLEO+OPE & Experiment\\\hline
$B\left(  \pi^{0}\rightarrow e^{+}e^{-}\right)  \times10^{8}$ & $\geq4.69$ &
$\geq5.85\pm0.03$ & $6.23\pm0.09$ & $7.49\pm0.38$ \cite{Abouzaid:2007md}%
\\\hline
$B\left(  \eta\rightarrow\mu^{+}\mu^{-}\right)  \times10^{6}$ & $\geq4.36$ &
$\leq6.23\pm0.12$ & $5.11\pm0.20$ & $5.8\pm0.8$ \cite{Abegg:1994wx}%
\\\hline
$B\left(  \eta\rightarrow e^{+}e^{-}\right)  \times10^{9}$ & $\geq1.78$ &
$\geq4.33\pm0.02$ & $4.60\pm0.06$ & $\leq2.7 \times10^4 \cite{Berlowski:2008zz}$\\\hline
\end{tabular}
\label{table2}%
\end{table*}

\section{Possible explanations of the effect}

Therefore, it is
extremely important to trace possible sources of the discrepancy between the
KTeV experiment and theory. There are a few possibilities: (1) problems with
(statistic) experiment procession, (2) inclusion of QED radiation corrections
by KTeV is wrong, (3) unaccounted mass corrections are important, and (4)
effects of new physics. At the moment, the last possibilities were
reinvestigated. In \cite{Dorokhov:2008qn}, the contribution of QED radiative
corrections to the $\pi^{0}\rightarrow e^{+}e^{-}$ decay, which must be taken
into account when comparing the theoretical prediction (\ref{Bt}) with the
experimental result (\ref{KTeV}), was revised. Comparing with earlier
calculations \cite{Bergstrom:1982wk}, the main progress is in the detailed
consideration of the $\gamma^{\ast}\gamma^{\ast}\rightarrow e^{+}e^{-}$
subprocess and revealing of dynamics of large and small distances.
Occasionally, this number agrees well with the earlier prediction based on
calculations \cite{Bergstrom:1982wk} and, thus, the KTeV analysis of radiative
corrections is confirmed. In \cite{Dorokhov:2008cd} it was shown that the mass
corrections are under control and do not resolve the problem. So our main
conclusion is that the inclusion of radiative and mass corrections is unable
to reduce the discrepancy between the theoretical prediction for the decay
rate (\ref{Bt}) and experimental result (\ref{KTeV}).

\section{$\pi^{0}\rightarrow e^{+}e^{-}$ decay as a filtering process for low mass dark matter}

\begin{figure}[th]
\includegraphics[width=7cm]{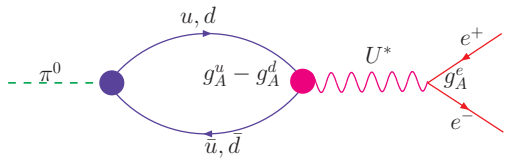}\caption{Loop diagram for $\pi
^{0}\rightarrow e^{+}e^{-}$ process induced by the low mass exotic $U^*$ boson.}%
\label{DM}%
\end{figure}

If one thinks about an extension of the Standard Model in terms of heavy, of an order of
$100$ GeV or higher, particles, then the contribution of this sort of particles to the
pion decay is negligible. However, there is a class of models for description of Dark
Matter with a mass spectrum of particles of an order of 10 MeV \cite{Boehm:2003hm}.
This model postulates a neutral scalar dark matter particle $\chi$
which annihilates to produce electron/positron pairs: $\chi\chi \to e^+e^-$. The excess
positrons produced in this annihilation reaction could be responsible for the bright 511
keV line emanating from the center of the galaxy \cite{Weidenspointner:2006nua}. The
effects of low mass vector boson $U$ appearing in such model of dark matter (Fig.
\ref{DM}) were considered in \cite{Kahn:2007ru} where the excess of KTeV data over theory
put the constraint on coupling which is consistent with that coming from the muon
anomalous magnetic moment and relic radiation \cite{FayetLatest}. Thus, the pion decay
might be a filtering process for light dark matter particles.

Further independent
experiments at KLOE, NA48, WASAatCOSY, BES III and other facilities will be crucial for
resolution of the problem. Also important is to get more precise data on the pion
transition form factor in asymmetric as well in symmetric kinematics.

\section{Acknowledgments}

We are grateful to the Organizers for a nice meeting and kind invitation
to present our results. Discussions on the subject of this work
with  C. Bloise, G. Collangelo, M.A. Ivanov, A. Kupcsh, E.A. Kuraev,
M. P. Lombardo, E. Tomasi-Gustafsson were very helpful.

\end{document}